\documentclass[10pt, aps, prb, twocolumn,
longbibliography, superscriptaddress]{revtex4-2}
\usepackage{soul}
\usepackage[utf8]{inputenc}
\usepackage[T1]{fontenc}
\usepackage{ragged2e}
\justifying
\usepackage{orcidlink}
\usepackage{amssymb}
\usepackage{amsmath}
\usepackage{graphicx}
\usepackage{bm}
\usepackage{xcolor}
\usepackage{enumitem}
\usepackage{array}
\usepackage{hyperref}
\usepackage{listings}
\usepackage{comment}
\usepackage{setspace}
\usepackage{datetime}
\usepackage{booktabs}
\usepackage{mathtools}
\usepackage{slashed}
\usepackage[normalem]{ulem}

\newcommand{\be}{\begin{equation}}
\newcommand{\ee}{\end{equation}}

\newcommand{\ra}{\rangle}
\newcommand{\p}{\partial}

\begin{document}

\title{Quantum Entanglement of Anyonic Charges and Emergent Spacetime Geometry}

\author{Hoang-Anh Le\,\orcidlink{0000-0002-1668-8984}}
\thanks{These authors contributed equally to this work.}
\affiliation{%
Center for Quantum Nanoscience, Institute for Basic Science, Seoul 03760, Republic of Korea
}%
\affiliation{%
Ewha Womans University, Seoul 03760, Republic of Korea
}%

\author{Hyun Cheol Lee\,\orcidlink{0000-0003-1352-8851}}
\thanks{These authors contributed equally to this work.}
\affiliation{%
Department of Physics, Sogang University, Seoul 04107, Republic of Korea
}%

\author{S.-R. Eric Yang\,\orcidlink{0000-0003-3377-1859}}
\thanks{Corresponding author: eyang812@gmail.com}
\affiliation{%
Department of Physics, Korea University, Seoul 02841, Republic of Korea
}%

% \email{eyang812@gmail.com}

%TC:ignore
\begin{abstract}
Intrinsically topologically ordered phases can host anyons. Here, we take the view that entanglement between anyons can give rise to an emergent geometry resembling Anti-de Sitter (AdS) space. We analyze the entanglement structure of fractionalized anyons using mutual information and interpret the results within this emergent geometric framework. As a concrete example, we consider pairs of 
$e/2$-charged semions that arise from instanton configurations in a disordered zigzag graphene nanoribbon. These fractional charges, located on opposite zigzag edges, show long-range quantum entanglement despite being spatially separated. We analyze the scale dependence of their entanglement and embed the ribbon into an AdS-like bulk geometry. In this setup, the entanglement structure defines minimal surfaces in the bulk, providing a geometric view of the edge correlations. This gives a holographic picture of fractionalized degrees of freedom in quasi-one-dimensional systems and shows how quantum entanglement can generate emergent geometry even without conformal symmetry.
\end{abstract}
%TC:endignore

\maketitle

\section{Introduction}

Intrinsically topologically ordered phases~\cite{wen2017,HaldaneNobel,yangbook} with topological entanglement entropy {\color{black}(TEE)}~\cite{Wen2006,Kitaev2006} support anyons~\cite{Laughlin1983}, which may carry fractional electric charge or remain electrically neutral~\cite{Feldman2021}. When an electron tunnels into such a system, it can fractionalize into multiple anyons, each obeying nontrivial quantum statistics~\cite{Haldane1991} that are neither fermionic nor bosonic. The stability of fractional anyon charges is inherently linked to suppressed fluctuations in the value of the TEE~\cite{Le_2024}.

In this work, we focus on a different facet of anyons: their mutual entanglement. Although fractional charges are often treated as independent particles, they are in fact quantum mechanically entangled.   
The mutual information between such particles is nonzero, even when they are far apart.  It reflects non-local correlations in the ground state and serves as a signature of quantum entanglement.  

The holographic principle~\cite{Hooft1993, Susskind1995} posits that all physical information in a volume can be encoded on its boundary, suggesting that spatial connectivity emerges from quantum correlations. This idea is realized concretely in Anti-de Sitter (AdS) space, where entanglement gives rise to an emergent bulk geometry, i.e., an information-theoretic geometry~\cite{Maldacena1997, Witten1998,Ryu2006}. Van Raamsdonk~\cite{van2010} further proposed that connected classical spacetimes require sufficient quantum entanglement; reducing this entanglement leads to disconnected geometries, implying that entanglement acts as the “glue” binding spacetime.

\textcolor{black}{This insight motivated Swingle’s proposal~\cite{Swingle2012} that tensor networks such as the multi-scale entanglement renormalization ansatz (MERA)~\cite{Vidal2007,orus2014} provide a discrete realization of AdS geometry through entanglement renormalization.}  MERA uses unitary disentanglers between coarse-graining steps to remove short-range entanglement while preserving long-range correlations, enabling efficient representations of critical states. The hierarchical layers of the network reflect correlations on different length scales, leading to a hyperbolic geometry. Although entanglement-induced geometry does not always need to resemble AdS, AdS provides a particularly transparent and well-studied example.

In parallel, Maldacena and Susskind introduced the ER = EPR conjecture~\cite{Maldacena2013}, identifying entanglement (EPR pairs~\cite{Einstein1935}) with spacetime connections (Einstein-Rosen bridges~\cite{Einstein1935_2}). In AdS/CFT, this is realized as entangled black holes on the boundary that are connected by wormholes in the bulk. Crucially, ER = EPR does not depend on conformal symmetry, it extends to a wide class of quantum systems, including non-conformal ones. Even in such cases, as in MERA, the entanglement structure retains a geometric character, supporting the broader principle that spacetime connectivity arises from quantum entanglement.

\textcolor{black}{Interestingly,  Einstein’s equations follow from the idea that the vacuum state of quantum fields has maximal entanglement in small regions of spacetime~\cite{Jacobson2016}. When matter or geometry is slightly perturbed, spacetime responds in such a way that the total entanglement entropy remains unchanged. This requirement uniquely leads to Einstein’s field equations.}

Connections between entanglement and emergent geometry have also been extensively explored in condensed matter systems. Near quantum critical points, entanglement scaling reveals intriguing patterns reminiscent of holographic duality~\cite{Vidal2003,sachdev2011,sachdev2012}. The variance of TEE across topological phase transitions is reflected in the magnitude of the spatial variation of the mutual information~\cite{mut_inf2023}.
Broader applications of string-theoretic methods in condensed matter physics are reviewed in Refs.~\cite{Hartnoll2009,Nastase2017book}.

Disordered zigzag graphene nanoribbons~\cite{jeong2019,yang2020,yang2021} (ZGNRs) and two-leg electron ladders~\cite{Yang2025} exhibit robust long-range entanglement, characterized by a \emph{universal} value of the TEE that remains invariant under variations in interaction strength and disorder~\cite{yang2021}—a hallmark of intrinsic topological order. Both systems host semions charged with \(e/2\) with statistical angle \(\pi/2\) and belong to the same universality class~\cite{Yang2025}. \textcolor{black}{In these phases, the \textit{ground state} already encodes localized semionic degrees of freedom, in contrast to other intrinsically topological phases where anyons appear only as excitations.}  Unlike two-leg electron ladders, the entangled fractional charges in ZGNRs are well separated over many lattice spacings, enabling the definition of an emergent space (information geometry) from the mutual information.
 Similar entanglement structures are expected in other systems that support anyonic excitations.

We find that entanglement between fractional charges in ZGNRs acts as a form of ``glue'', giving rise to an emergent AdS-like geometry that effectively connects the two edges through entanglement patterns resembling geodesics, reflecting the spirit of the holographic principle.  Mutual information quantifies non-local correlations and encodes effective distances through geodesic lengths or, in higher dimensions, minimal surface areas. (A geodesic minimizes length, while a minimal surface minimizes area.) This picture echoes the ER = EPR proposal, suggesting that entanglement defines geometry, even in non-CFT systems like ZGNRs, where scale-dependent entanglement supports emergent spatial structure beyond the AdS/CFT paradigm.

This paper is organized as follows.
In Section~\ref{sect:instantons} we explain the types of non-local correlations that emerge in a zigzag graphene nanoribbon (ZGNR) hosting a pair of fractional charges localized on opposite zigzag edges.
In Section~\ref{ScaleMI}, we analyze the mutual information between these edge-localized charges, with particular emphasis on its scale dependence, which is illustrated graphically.
In Section~\ref{embedding_hyperbolic}, we conceptually embed the ZGNR in a curved space, treating it as if it resides in a non-Euclidean hyperbolic-like geometry. 
In Section~\ref{emergent_geodesic}, we introduce the geodesic distance in this emergent geometry, inferred from the coarse-grained mutual information. 
Finally, Section~\ref{summary} presents a summary of our work.

\section{Instantons and the Emergence of Topological Charge in ZGNRs}
\label{sect:instantons}

Disordered ZGNRs support topological fractional edge charges. In the clean and non-interacting limit, they behave as a symmetry-protected topological insulator. However, on-site electron interactions break chiral symmetry, allowing a Goldstone–Wilczek-type mechanism to generate a topological charge of one, rather than one-half. In the presence of disorder, these topological states become coupled, giving rise to instantons that carry a fractional charge \( e/2 \).  Atomically precise ZGNRs have been fabricated in the last ten years~\cite{Ruffieux2016,Kolmer2020,houtsma2021atomically}.

The following subsections~\ref{subsect:sec1}, \ref{sec2}, \ref{sec3} establish the notation and review key results previously obtained for intrinsically topologically ordered ZGNRs, thereby laying the foundation for our analysis of the spatial structure of 
entanglement between $e/2$ semionic charges.

\subsection{Hamiltonian}
\label{subsect:sec1}

Disordered ZGNRs are modeled using a Hubbard Hamiltonian that includes the interaction term with on-site repulsion \( U \) at each site and the disorder term implemented through random on-site potentials $V_m$. In the Hartree-Fock mean-field approximation, the Hamiltonian takes the following form
\begin{equation}
\begin{aligned}
H_{\text{MF}} = &-t\sum_{\langle m,n \rangle,\sigma} \widehat{c}^{\dag}_{m,\sigma} \widehat{c}_{n,\sigma} +
\sum_{m,\sigma} V_m \widehat{c}_{m,\sigma}^{\dag} \widehat{c}_{m,\sigma}  \\
&+U \sum_m [ \widehat{n}_{m,\uparrow}\langle \widehat{n}_{m,\downarrow}\rangle +\widehat{n}_{m,\downarrow} \langle \widehat{n}_{m,\uparrow}\rangle-\langle \widehat{n}_{m,\downarrow}\rangle \langle \widehat{n}_{m,\uparrow}\rangle].
\label{MFhspin}
\end{aligned}
\end{equation}
Here, \( t \) denotes the hopping amplitude, \( \widehat{c}^\dagger_{m, \sigma} \) the fermion creation operator at site \( m \) with spin \( \sigma \), and \( \widehat{n}_{m,\sigma} \) the corresponding number operator. The site index \( m \) refers to the positions and the spin index takes values \( \sigma = \uparrow \) (spin-up) or \( \downarrow \) (spin-down).
The sum of $\langle m, n \rangle$ specifies that the hopping is between the nearest neighbors. 
The second term represents a short-range disorder potential. %with 10\% of the total site of the ribbon that contains impurities. 
The magnitude of the potential at each site $V_m$ is drawn from a uniform distribution $[-\Gamma, \Gamma]$. In the continuum limit, the strength of the disorder is characterized by \( n_{\mathrm{imp}} \Gamma^2 \), where \( n_{\mathrm{imp}} \) denotes the density of impurities. In the following, we set \( n_{\mathrm{imp}} = 10\% \) and our results are not sensitive to the precise value of \( n_{\mathrm{imp}} \)~\cite{yang2020}.
The Hartree–Fock ground state agrees well with the results obtained from the formulation of the matrix product state of the density matrix renormalization group (DMRG)~\cite{yang2022} and from bosonization~\cite{Yang2025}. The presence of a gap and localization effects suppresses quantum fluctuations, making the Hartree–Fock approximation particularly effective. In this work, we adopt the Hartree–Fock approach, which enables efficient exploration of large systems.

% \pagebreak

% \begin{figure}[ht!]
% \centering
% \includegraphics[width=0.4\linewidth]{figure1.pdf}
% \caption{Each carbon site is labeled by \( m=(h, k) \), where \( h \) denotes the row index and \( k \) the column index. Note that each site labeled by $k$ may correspond to different physical positions across horizontal carbon lines, due to their relative displacement within the graphene lattice.  Red (blue) sites are labeled by chirality A (B). The opposite zigzag edge sites that form upper and lower boundaries have different chiralities.}
% \label{partition}
% \end{figure}

\subsection{Breaking of Chiral Symmetry and Goldstone-Wilczek Approach}\label{sec2}

It is instructive to first consider the disorder-free zigzag ribbon.
In the absence of disorder, the non-interacting ZGNR is a symmetry-protected topological (SPT) insulator in the sense that the gapless edge states exist owing to a sublattice symmetry called \textit{chiral} symmetry.
The chiral transformation is defined as
\be
\label{chiral_sublattice}
\widehat{\mathsf{a}}_i  \to +\widehat{\mathsf{a}}_i, \quad 
\widehat{\mathsf{b}}_i  \to -\widehat{\mathsf{b}}_i,
\ee
where \( \widehat{\mathsf{a}}_i \) and \( \widehat{\mathsf{b}}_i \) are the electron annihilation operators on the A and B sublattices, respectively. (The operators \( \widehat{c}_i \) appearing in the Hamiltonian Eq.~\eqref{MFhspin} can be grouped into two types, \( \widehat{\mathsf{a}}_i \) and \( \widehat{\mathsf{b}}_i \).)  Under the chiral transformation, the non-interacting Hamiltonian (the first term of Eq.~\eqref{MFhspin}) changes sign.

\textcolor{black}{Symmetries can be analyzed more systematically using a low-energy effective theory. In this regime, the continuum limit of the Hubbard model of ZGNR is described by a Dirac Hamiltonian.}  The axial symmetry transformation of massless Dirac fermions in \textit{odd} spatial dimensions is
\be
\label{twoglobal}
\psi \to e^{i \Lambda} \psi \quad \text{(ordinary)}, \quad
\psi \to e^{i \gamma^5 \Lambda} \psi \quad \text{(axial or chiral)},
\ee
where $\psi$ is a Dirac spinor and $\Lambda$ is a global phase. In one spatial dimension, $\gamma^5 = \gamma^0 \gamma^1$.
Chirality is defined as the eigenvalue of $\gamma^5$, which can take either the $+1$ ($R$ component) or the $-1$ ($L$ component). According to \textit{classical} Noether theorem, the invariance of 
Lagrangian under the transformations of  Eq.~(\ref{twoglobal}) implies the conservation of the corresponding 
currents ($\gamma^\mu$'s are Dirac matrices and $\mu =0$ is the time component):
\be
j^\mu_{\rm ordinary} = \bar{\psi} \gamma^\mu \psi, \quad 
j^\mu_{\rm axial} = \bar{\psi} \gamma^\mu \gamma^5 \psi.
\ee
However, this axial symmetry is broken when a mass term is present.  A mass term couples $\psi_R$ and $\psi_L$, and the axial current is no longer conserved, and the continuity equation for the axial currents in one spatial dimension takes the following form
\cite{goldstone1981}:
\be
\label{anomaly-eq}
\p_\mu j^{\mu 5} =  \frac{\p j^1}{\p t} + \frac{\p j^0}{\p x} = \mathcal{M}(x),
\ee
where $ \mathcal{M}(x)$ is the mass term that may depend on coordinates.
Goldstone and Wilczek derived an expression for the quantum number of fractional fermions
in the background of the topologically nontrivial mass term $\mathcal{M}(x)$ (after subtracting a 
vacuum contribution) in terms of the \textit{topological charge} ($Q_{\rm top}$ below)~\cite{goldstone1981}.
The integration of Eq.~(\ref{anomaly-eq}) (in the temporal adiabatic limit) between boundary yields ($\rho(x) = j^0(x)$)
\be
\rho(\infty)-\rho(-\infty) = \int^\infty_{-\infty} dx \, \mathcal{M}(x) = Q_{\rm top}.
\label{eq:Qtop}
\ee
For polyacetylene $\mathcal{M}(x)$ is equal to the
spatial derivative of the kink dimerization order, so the boundary behavior of the dimerization
order parameter determines the anomalous fermion number.

Now, if the on-site electron repulsion is included, the chiral sublattice symmetry of Eq.~\eqref{chiral_sublattice} is \textit{explicitly} broken since the density
operator does not change under this transformation.
We may expect that the edge states of noninteracting ZGNR may be gone. However, it turns out that
the edge states persist, but now they are \textit{gapped}. %~\cite{chiral-ZGNR}
These gapped edge states can be understood from the point of view of the relation between
the fermion number and the topological configuration of the gap parameters analogous to the Goldstone-Wilczek method discussed above~\cite{chiral-ZGNR}. 
\textcolor{black}{The combined mirror and time-reversal symmetry is essential for determining the system’s topological properties and for maintaining its SPT classification.}
More concretely, one can construct the analog of anomaly equation as in Eq.~\eqref{anomaly-eq}
for the Hartree-Fock Hamiltonian by considering the time derivative of (total) spatial current
using the Heisenberg equation of motion.  A careful study of the equation for the current perpendicular to edges shows that the edge occupations of the ZGNR can be identified with the topological charge
linked to the behavior of the gap parameters. 

\textcolor{black}{The arguments leading to the topologicla charge $Q_\text{top}$ in Eq.~\eqref{eq:Qtop}  apply only to symmetry-protected topological insulators (SPTs), which in our case correspond to disorder-free ZGNRs. In this regime, semions are absent: while edge states exist, they carry integer charge $e$, not fractional charge $e/2$ as in polyacetylene~\cite{HeegerMod1988}. A detailed discussion of these differences can be found in Ref.~\cite{Yang2025,chiral-ZGNR,Yang2019/1}.}

It is tempting to try to understand the interacting ZGNR  in the scheme of the symmetry-based classification of topological insulators/superconductors as proposed in Refs.~\cite{tenfold1,tenfold2}.
However, it is not clear to what extent such a classification scheme can be applied since the scheme is based on the condition of the preservation of \textit{gapless
boundary states} against various perturbations that adhere to relevant symmetries.

\subsection{Role of Disorder}\label{sec3}

Up to this point, we have neglected disorder, which can strongly impact the system {\it beyond the symmetry-protected features}. \textcolor{black}{Disorder may originate from random defects and/or random variations in the hopping parameters. Such disorder breaks several symmetries, including chiral, mirror, rotational, and translational symmetries, and is not confined to the edges of the system. Disorder induces Anderson localization, an emergent property which requires a macroscopic number of impurities.  Nonlocal correlations extending across the entire 
ribbon~\cite{Le_2024} 
(to be discussed below) are present only when disorder is distributed throughout the ribbon. The interplay between disorder and electron-electron interactions is highly nontrivial, giving rise to rich and complex physical behavior.}

\begin{figure}[ht!]
    \centering
\includegraphics[width=0.8\linewidth]{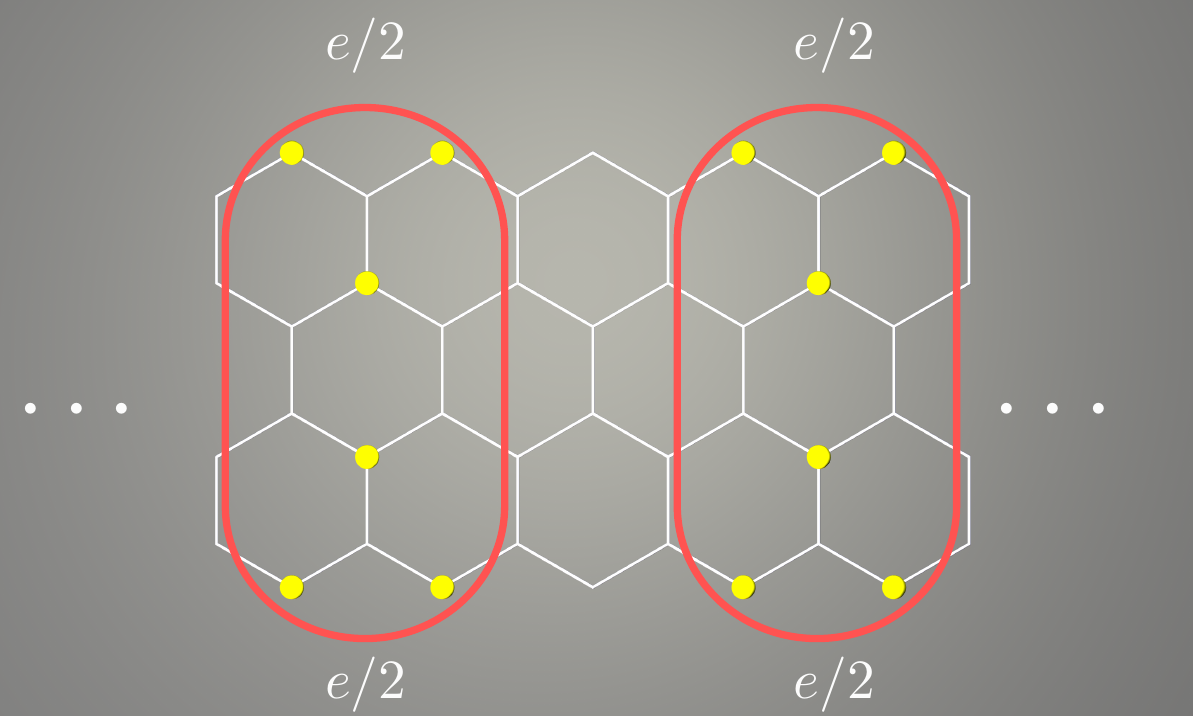}
     \caption{In the disordered \textit{ground} state of a zigzag ribbon, fractional charge pairs are localized at opposite edges. Yellow dots schematically represent the site-resolved probability density of a fractional charge. It decays rapidly from the edges. While each $e/2$ fractional charge obeys semionic statistics, the composite object formed by a pair, highlighted by the red curve, obeys Fermi statistics~\cite{Yang2025}.  The number of composite objects scales linearly with the ribbon length.}
     \label{FC}
\end{figure}

Impurities couple the upper and lower zigzag edge states, generating instantons composed of two fractional charges \( e/2 \), each localized on one of the edges. In the \textit{ground} state, numerous such pairs are present, with each \( e/2 \) charge confined to one edge and its partner residing on the opposite edge (see Fig.~\ref{FC}). Upon doping, additional midgap states appear, each carrying a fractional charge~\cite{yang2022}.
 As a result, disorder drives the system from a symmetry-protected topological insulator to an intrinsically topologically ordered phase characterized by a finite TEE~\cite{yang2021}.
In addition, bosonization method shows that fractional charges exhibit semionic statistics such that their fusion yields an electron with ordinary fermionic statistics~\cite{Yang2025}.

An edge fractional charge generated by a spin-up instanton can overlap spatially with that from a spin-down instanton, forming a composite with unit electric charge and total spin zero~\cite{yang2020}. Such composites are more energetically favorable than individual fractional charges and may give rise to unusual signatures in combined spin susceptibility and transport measurements~\cite{yang2022}, similar to those observed in polyacetylene~\cite{Chung1984}.

Several effects are crucial for understanding the robustness of fractional charges. Electron localization plays a significant role in stabilizing fractional charges. Upon weak doping,  midgap localized fractionalized states with $E\approx 0$ exist~\cite{yang2022}, but they do \textit{not} spatially overlap each other which is a hallmark of localization~\cite{Altshuler, GV2000}. Furthermore, fractionalized states are isolated from other states of higher energy by an exponential gap in the density of states. This implies that quantum fluctuations have a negligible effect on fractional charges. 
%The results of the Hartree-Fock approach are consistent with those obtained from DMRG method~\cite{yang2022} and from bosonization analysis~\cite{yang2025}.
As the strength of disorder and/or interchain coupling deviates from the universal region in the parameter space, the variance of TEE is expected to grow, which results in crossover phases of quasi-topological order~\cite{Le_2024}. In this case, fractional charges are not well defined.

\subsection{Non-Local Correlations}

\begin{figure*}[ht!]
\centering   
\includegraphics[width=0.9\linewidth]{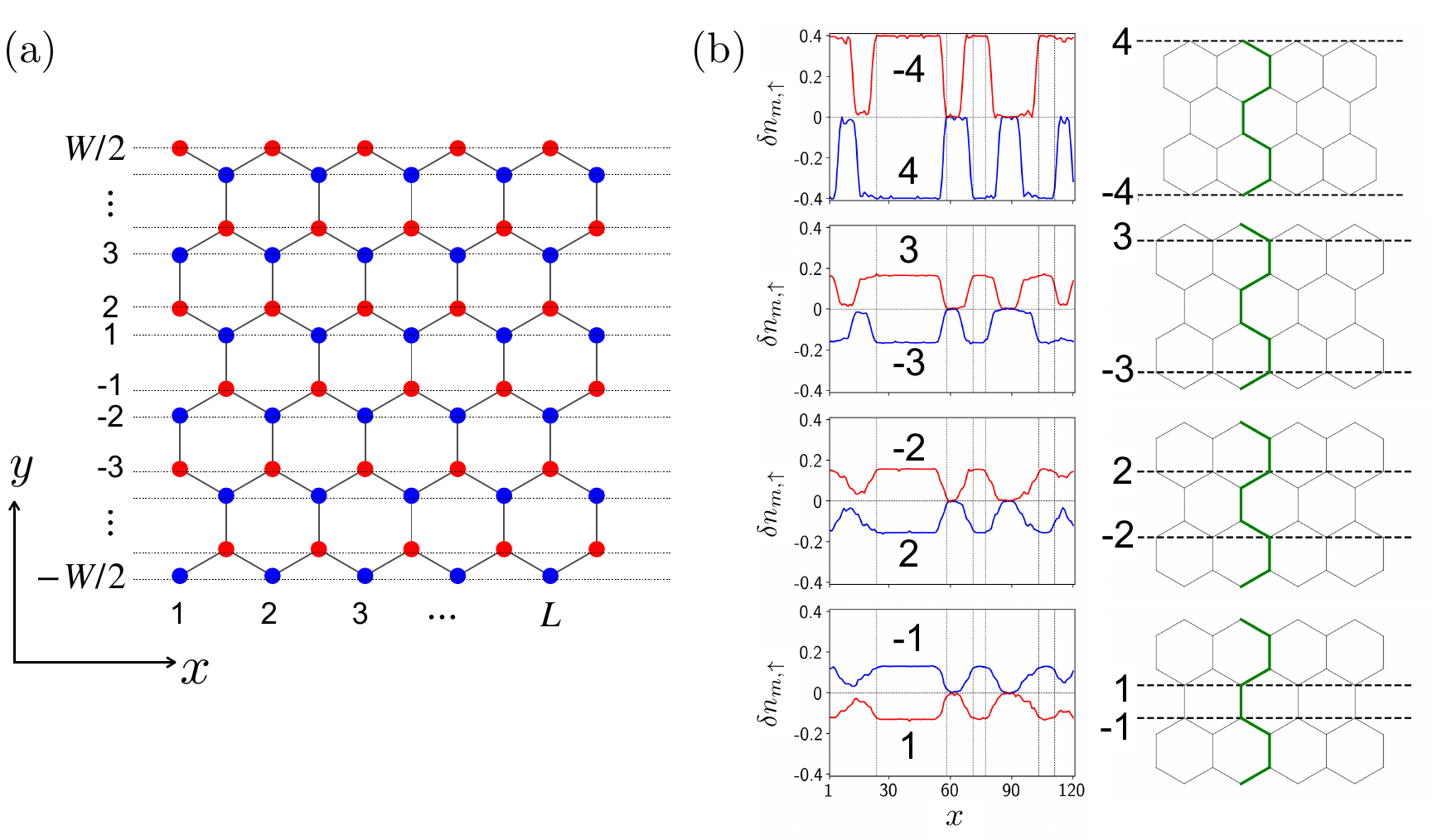}
\caption{(a) Site index for a ZGNR with length $L$ and width $W$. Each carbon site is labeled by \( m=(x, y) \), where \( x \) denotes the row index and \( y \) the column index. Note that each site labeled by $x$ may correspond to different physical positions across horizontal carbon lines, due to their relative displacement within the graphene lattice.  Red (blue) sites are labeled by chirality A (B). The opposite zigzag edge sites that form upper and lower boundaries have different chiralities.
(b) \textbf{Left:} Each panel shows \( \delta n_{m, \uparrow} \), the change in spin-up occupation at site \( m \) between disordered and clean cases, plotted along two opposite horizontal carbon chains. 
%The labels \( y = \pm 1, \pm2, \ldots, \pm 4 \) indicate the horizontal carbon chains (see the right panel).  
\textbf{Right:} Pairs of opposite chains are shown as horizontal dashed lines. The vertical green line corresponds to a vertical dotted line in the left panels; that is, all sites along the green line share the same horizontal index \( x \). Fractional charges appear near the two ends of the green line, where \( \delta n_{m, \uparrow} \) changes sharply. The Hamiltonian's parameters are \( (U, \Gamma) = (2t, 0.1t) \) and the ZGNR geometry is specified by \( (L, W) = (120, 8) \). {\color{black} The data was obtained from a single disorder realization.}}
\label{variation}
\end{figure*}

Fractional charges are intrinsically linked to non-local correlations. Although the charges themselves are edge localized, their correlations extend into the bulk along the axis joining the two charges (see  Fig.~\ref{variation}), as shown in the variation of the number of spin-up occupations, \( \delta n_{m, \uparrow}=n_{m, \uparrow}-n^0_{m, \uparrow} \), along horizontal carbon lines~\cite{Le_2024}. The quantities $n_{m,\uparrow}$ and $n^0_{m,\uparrow}$ represent the occupation numbers in the presence and absence of disorder, respectively.  Abrupt changes in \( \delta n_{m,\uparrow} \) are observed when a pair of fractional charges are present at the zigzag edges. Dotted vertical lines in the left panels of  Fig.~\ref{variation} (b) indicate the site indices \( x \) where these discontinuities occur. Remarkably, sharp variations at the same horizontal positions persist deep in the bulk, far from the physical edges. This behavior signals the presence of non-local correlations, which are observed similarly in the spin-down occupations \( \delta n_{m,\downarrow} \)~\cite{Le_2024}.
\textcolor{black}{Nonlocal correlations extending across the entire ribbon (see Fig.~\ref{variation} (b)) are present only when disorder is distributed throughout the ribbon. } These correlations are the origin of the scale-dependent mutual information discussed below.  
% in Section~\ref{scale_dependent_entanglement}. 

%In this context, the separation of a pair of horizontal carbon lines from the system's edges
%(see the right panels in Fig.~\ref{variation} (b)) may be viewed as a renormalization scale, analogous to the depth coordinate in MERA-like tensor network architectures.

% \pagebreak

\section{Entanglement Structure via Mutual Information in a Disordered Potential}
\label{ScaleMI}

An instanton consisting of two fractional charges \( e/2 \), each residing on an opposite edge, gives rise to long-range entanglement. In this section, we investigate the spatial structure of this entanglement via  mutual information.

\subsection{Mutual Information and von Neumann Entropy}

The entanglement entropy of a region \( D \) is defined as $S_D = -\text{Tr}[\rho_D \ln \rho_D]$, where \( \rho_D \) is the reduced density matrix of \( D\).
The entanglement entropy is computed as 
\begin{eqnarray}
S_D=-\sum_i
[\lambda_i \ln \lambda_i-(1-\lambda_i)\ln (1-\lambda_i)],
\end{eqnarray}
where $\lambda_i$ are the eigenvalues of the following correlation function (or equivalently, of the reduced density matrix of region $D$)~\cite{yang2021}
\begin{eqnarray}
C_{\vec{R},\vec{R}'}=\langle \Psi| \widehat{c}^\dagger_{\vec{R}\uparrow} \widehat{c}_{\vec{R}'\uparrow}|\Psi\rangle.
\label{Corr}
\end{eqnarray}
In Eq.~\eqref{Corr}, $\vec{R}, \vec{R}'\in D$ denote site positions restricted in region $D$ and $\vert \Psi \ra$ is the Hartree-Fock ground state.
This expression remains valid when applied to the correlation function of spin-down states as well.  Open boundary conditions are used. 

\textcolor{black}{The entanglement entropy of a spatial region $D$ contains a universal subleading
constant term $\gamma$, known as the  TEE
\cite{Wen2006, Kitaev2006}:
\begin{equation}
S_D = \alpha L - \gamma,
\end{equation}
where $L$ denotes the length of the boundary of the region $D$. The coefficient \( \gamma \) is a universal constant, independent of microscopic details, and characterizes the topological order in a certain region of the parameter space.  The TEE is a topological invariant in the graphene ribbon system. Disorder must be present both at the edges and in the bulk; otherwise, one does not obtain a universal value of the TEE that is independent of the shape of any smooth domain within the ribbon. This is a  stringent requirement.
The region $D$ used to compute the TEE must be chosen sufficiently far from the edges in order to suppress non-universal edge contributions and reliably extract the topological term~\cite{yang2021}. 
}

\textcolor{black}{Defining the von Neumann entanglement entropy for identical fermions is generally subtle, due to the non-factorizable structure of the many-body Fock space and the resulting ambiguity in partial traces (see, e.g., Ref.~\cite{Balachandran2013}). In the present work, however, this issue does not arise because we consider spatial bipartitions on a lattice, for which the fermionic Fock space factorizes into a tensor product of local mode algebras. The ambiguity occurs only for particle-based partitions, whereas for spatial bipartitions the reduced density matrix and entanglement entropy are well defined \cite{Zanardi2002, Wiseman2003}.}

In quantum information theory, mutual information quantifies the total correlations between two separate regions of a quantum system. It is defined in terms of entanglement entropy and is particularly well suited for probing non-local entanglement between spatially separated fractional charges, such as those localized on opposite zigzag edges.
We compute the Hartree-Fock mutual information~\cite{mut_inf2023} between two regions $I$ and $J$, defined as
	\begin{eqnarray}
	M_{I,J}=S_{I}+S_{J}-S_{I \cup J}.
\end{eqnarray}
Here, $S_{I}$, $S_{J}$, and $S_{I \cup J}$ denote the entanglement entropy of regions $I$, $J$, and their union $I \cup J$, respectively.  We define the region \( I \) (\( J \)) as sub-regions.
%; see Fig.~\ref{partition2}.  
Mutual information depends not only on the distance between the regions \( I \) and \( J \), but also on their sizes, reflecting the  scale-dependent nature of entanglement. 
\textcolor{black}{The definition of the mutual information is valid for both SPT phases and intrinsically topologically ordered phases, independent of whether disorder is present.}

\subsection{Mutual Information of an Instanton}

\begin{figure*}[ht!]
    \centering
    \includegraphics[width=\linewidth]{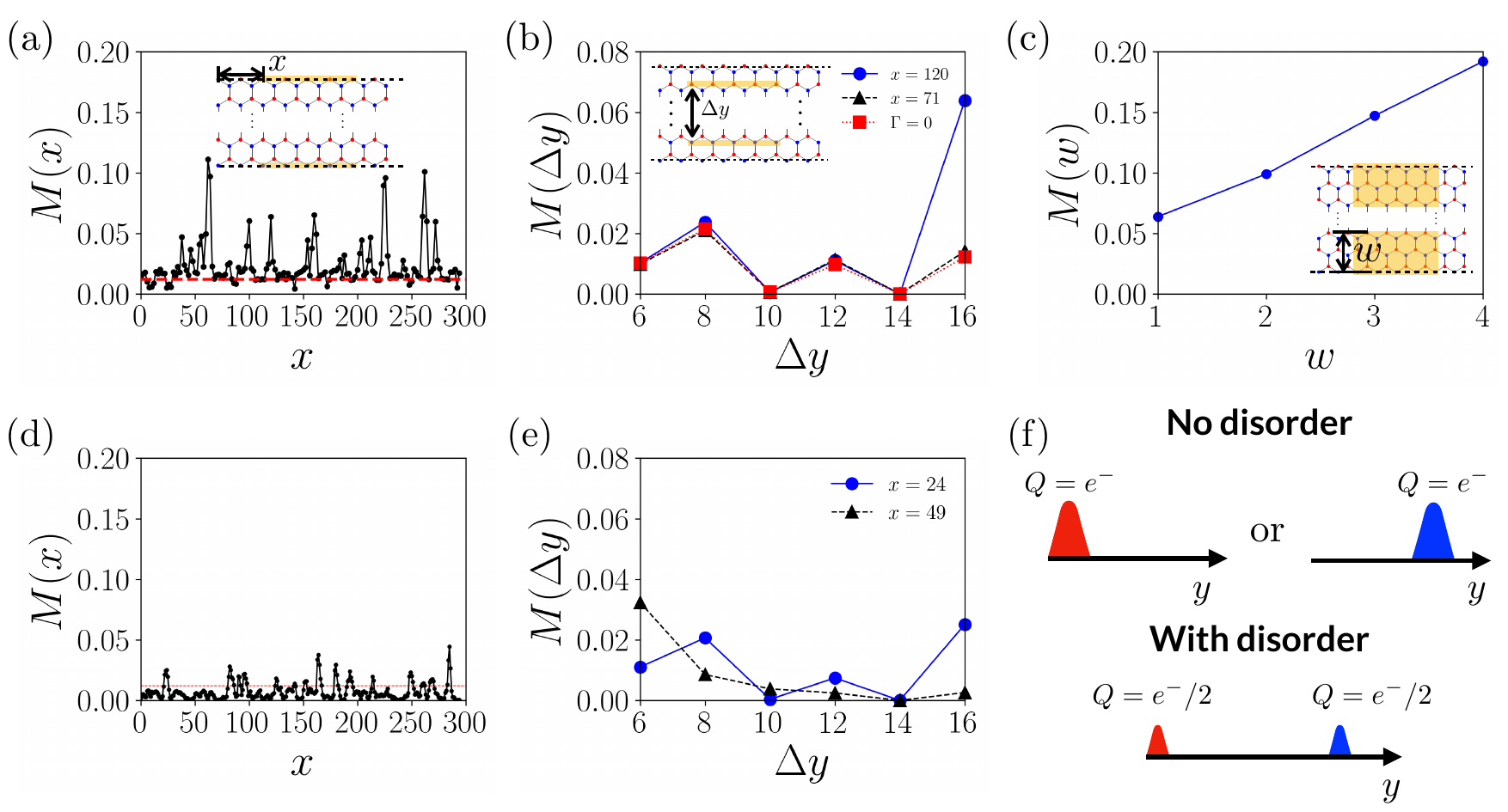}
    \caption{ {\color{black}{(a) Mutual information between two yellow boxes as a function of their position $x$, measured from the left end (see inset), obtained from a single disorder realization. Each box has size $(\ell, w) = (5, 1)$, where $\ell$ is the length along a zigzag edge and $w$ is the width measured in carbon rows.  
Here $(U, \Gamma) = (t, 0.01t)$. The red horizontal line indicates $M(x)$ of the translationally invariant, disorder-free case $(\Gamma = 0)$. Horizontal dashed lines in the insets indicate zigzag edges.
(b) Mutual information  at $x=120$ (hosting a fractional charge) and $x=71$ (without a fractional charge), plotted as a function of the vertical separation $\Delta y$ between the two boxes, with a maximum separation of $\Delta y = W = 16$.   Here $(U, \Gamma) = (t, 0.01t)$.  For comparison, we also show the mutual information of  the disorder-free system with $(U,\Gamma) = (t,0)$.   (c) Mutual information as a function of box width $w$, with boxes positioned to enclose a localized fractional charge at $x = 120$. (d) Mutual information versus position $x$ in the strong-disorder regime $(U, \Gamma) = (t, 12t)$, obtained from a single disorder realization. (e) Mutual information plotted as a function of \( \Delta y \) at two selected positions in panel (d). These curves are representative of the \emph{typical} behavior at various values of \( x \).
    (f) Schematic illustration of the probability density of edge states with and without disorder. In the absence of disorder, the single-particle state is localized on either zigzag edge at $y = \pm W/2$ (perpendicular to the zigzag edge direction), while remaining delocalized along the zigzag edge. In the presence of weak disorder, the single-particle edge state becomes fractionalized into two spatially separated halves, each localized on one of the two edges.  This state is localized both perpendicular to and along the zigzag direction. All numerical calculations are performed for a ribbon with ($L,W)=(300,16)$.}}}
    \label{fullMI}
\end{figure*}

We now investigate how non-local correlations are reflected in the mutual information. 
Figure~\ref{fullMI}(a) shows, in the weak disorder regime, the mutual information between two boxes, each consisting of five sites along the zigzag edge, plotted as a function of their distance \( x \) from the left end of the ribbon.  \textcolor{black}{The box size is chosen to optimally reveal the behavior of \( M \) as a function of \( x \) and \( y \), while reflecting the fact that semions are localized on the zigzag edges, with their probability density decaying into the bulk within a few lattice constants.  The horizontal and vertical dimensions of each box are chosen to be comparable to the size of the semions, which sets the relevant length scales of the problem.}  
Whenever a box contains a fractional charge \( e/2 \), corresponding to a site where the occupation number \( \delta n_{m,\uparrow} \) changes abruptly (see Fig.~\ref{variation}(b)), the mutual information exhibits a pronounced peak as a function of \( x \). Because fractional charges are localized in random positions, the overall profile exhibits a sequence of alternating peaks and valleys.

\textcolor{black}{Figure~\ref{fullMI}(b) shows examples of $M(\Delta y)$  vs $\Delta y$ at some selected values of $x$. 
They compare mutual information as a function of the vertical distance $\Delta y$ in two cases, depending on whether the horizontal box encloses a fractional charge or not.
In both cases, each box consists of five sites along a single row, i.e. $\ell = 5$ and \( w = 1 \). In the first case, the mutual information \( M \) remains small up to \(\Delta y = W-1\), where a pronounced peak emerges.  In contrast, in the second case, where the boxes do not contain fractional charges at the edges, no such peak is observed in \( M \) as \( \Delta y \) increases. Figure~\ref{fullMI}(b) also shows the disorder-free case, in which no peak is found either.  In other words, in the absence of semions, no peak appears 
at \(\Delta y = W-1\) and the edge wave functions do 
\textit{not} spatially split between opposite edges (see Fig.~\ref{fullMI}(f)).  
Figure~\ref{fullMI}(e)  displays
the mutual information $M(\Delta y)$ at two values of $x$ in the strong-disorder regime.  They do {\it not} exhibit a pronounced peak at the edges. The absence of semions in the strong-disorder regime is consistent with the observation that the gap in the density of states closes and the topological order is destroyed~\cite{Le_2024}.}

\subsection{Scale-Dependent Entanglement}
\label{scale_dependent_entanglement}

\begin{figure}[ht!]
\centering
\includegraphics[width=0.8\linewidth]{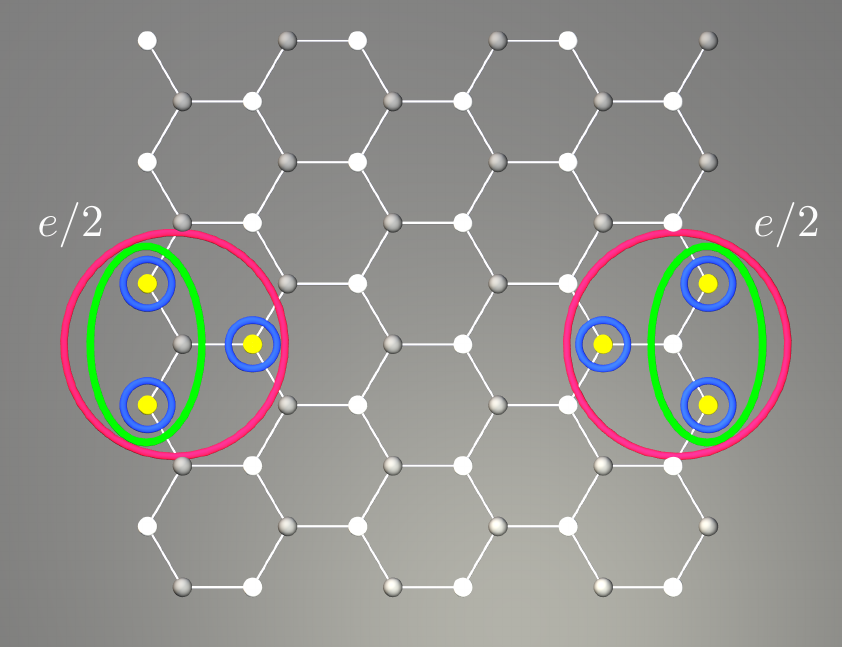}
\caption{Schematic illustration of the multiscale mutual entanglement pattern induced by a pair of fractional charges located on opposite zigzag edges.  Grey (white) dots represent carbon sites on the A (B) sublattice.  Each closed curve defines a region
%, labeled \( I \) or \( J \), 
used to compute the mutual information.
Each curve color corresponds to a scale of entanglement (e.g., blue = short range, red = longer range). When two closed curves of the same color appear near {\it opposite edges} of the ribbon, they enclose regions that are strongly entangled with each other. For instance, a pair of blue circles on opposing sides indicates entanglement at the shortest scale. The scale increases with the color gradient, progressing from blue to red. Only the strongest mutual information links are shown. Color legend: blue indicates 1 layer, green indicates 2 layers, and red indicates 3 layers. Links involving more than 3 layers are omitted for clarity.
Also, pairs of opposite sites with weak mutual information are not enclosed by curves.}
\label{fig:entanglement_pattern}
\end{figure}

We have also explored scale-dependent entanglement by computing mutual information for regions (boxes) that span multiple rows of carbon sites, as presented in Fig.~\ref{fullMI}(c). 
%The results show clear signatures of non-local long-range correlations, as shown in Fig.~\ref{variation}  [\textcolor{blue}{ I don't see a clear connection with  Fig.~\ref{variation}. Maybe need  more explanation: Hyun}]. 
Together with Figs.~\ref{fullMI}(a) and \ref{fullMI}(b), which present single-row configurations, these findings reveal the characteristic entanglement pattern surrounding a fractional charge \( e/2 \), schematically depicted in Fig.~\ref{fig:entanglement_pattern}.
This figure illustrates the scale-dependent entanglement structure associated with a fractionalized charge, where ``scale'' refers to the number of sites included in region \( I \) or \( J \). Note that short-range entanglement between \( I \) and \( J \) is excluded from the mutual information \( M_{I,J} \), since the two regions are spatially separated on opposite sides of the system.  The grouping of neighboring degrees of freedom into larger blocks allows one to probe entanglement at increasing length scales, a concept known as scale-dependent entanglement~\cite{Vidal2007}. The renormalization scale thereby defines a holographic direction along which bulk correlations decay exponentially~\cite{Swingle2012}.

 %\pagebreak

\section{Embedding of ZGNR in Hyperbolic-like Geometry}
\label{embedding_hyperbolic}

Although AdS spacetime does not directly emerge from the microscopic physics of condensed matter systems, a curved geometry (information geometry) with analogous features can arise through an embedding. 
A particularly relevant example is the \emph{soft-wall} geometry~\cite{Andreev2006, Karch2006, Klebanov2008, Erlich2014}, which introduces a smooth asymptotic boundary at \( y = \pm W/2 \) without imposing hard cutoffs, where $W$ is the ribbon's width. %(see Fig.~\ref{HPG}). 
Near these edges, the embedding generates a diverging metric that acts as an effective boundary, gradually suppressing the propagation of low-energy electrons toward \( y = \pm W/2 \).  A non-isometric embedding captures the underlying information structure, such as patterns of correlation or entanglement, rather than the actual spacetime metric structure. (That is, it does not preserve distances or local geometry.)

We construct an isometric embedding of a ZGNR into three-dimensional Euclidean space and show that geodesics on the resulting non-Euclidean surface closely resemble those in a hyperbolic-like geometry.
 We consider the system in the absence of disorder in the continuum limit of a quasi-one-dimensional ribbon, described by the coordinates $(x, y)$ as depicted in Fig.~\ref{variation}(a).
The pullback metric is  
\begin{equation}
ds^2 
=   R(y)^2 dx^2+dy^2,
\label{eq:induced_metric}
\end{equation}
The dimensionless radius of local curvature at depth \(y\) for our soft-wall geometry is given by
\begin{equation}
R(y) = \frac{W}{y_0 \cosh^2(y / y_0)},
\label{eq:Ry}
\end{equation}
where \(y_0\) sets both the central value \(R(0) = W / y_0\) and the characteristic decay scale along \(y\) axis.
The coordinate \( y \) is restricted to the interval \( |y| \le W/2 \).  The embedded ribbon has a negative Gaussian curvature (see Appendix~\ref{GCurv}).  
The pullback metric is used to induce a metric on the surface itself. This induced metric allows you to calculate the distances between two points on the embedded surface by pulling the ambient-space metric back on the surface. 

\begin{figure*}[ht!]
\begin{center}
\includegraphics[width=\textwidth]{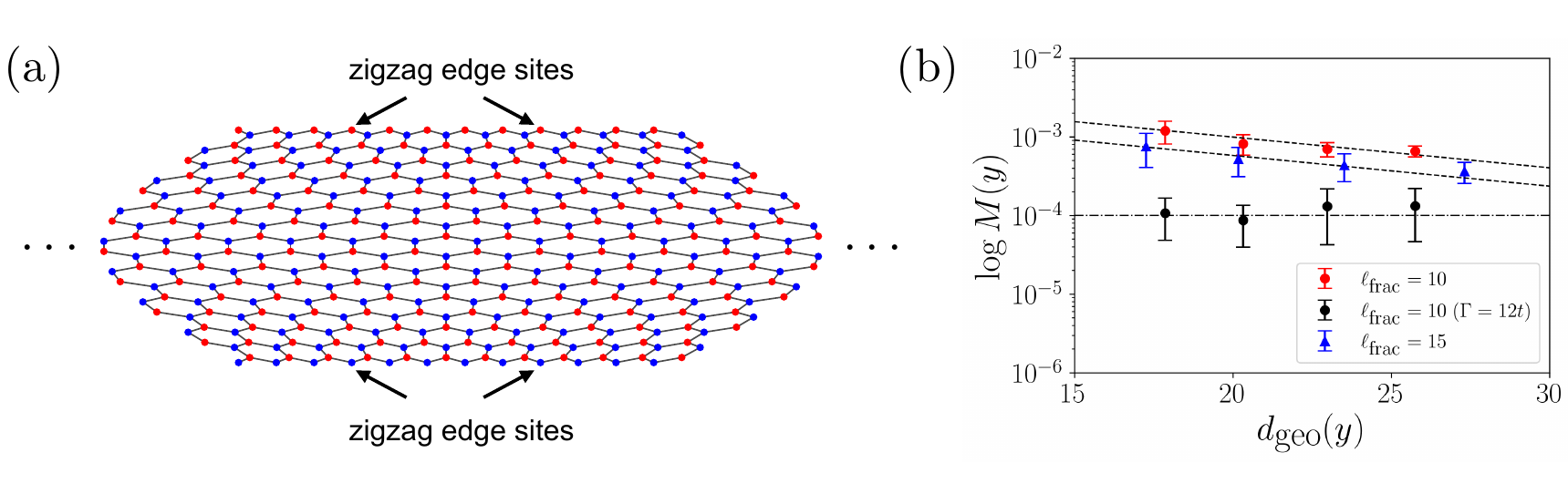}
\caption{(a) ``Hyperbolic geometry'' of a ZGNR with width \( W = 32 \) and characteristic curvature scale \( y_0 = 9 \).
A and B carbon sites are represented by red and blue colors, respectively. The ribbon comprises two interpenetrating triangular sublattices, shown in red and blue. The red and blue zigzag edge sites lie directly opposite each other across the ribbon. 
{\color{black} (b) The coarse-grained (disorder-averaged) mutual information 
$\log M(y)$ is computed with a transverse width $w = 1$, for typical separations between fractional charges $\ell_{\text{frac}} = 10,\,15$.  Larger $d_{\mathrm{geo}}(y)$ corresponds to points closer to the ribbon center.
The fitted slope (dashed lines) yields an inverse decay length $\alpha = 0.09$, with prefactors $A = 0.006$ for the red data points and $A = 0.0035$ for the blue data points (see Eq.~\eqref{eq:MI_decay}), with the curvature scale parameter fixed at $y_0 = 10$. 
The blue and red data points were computed with parameters in topological order regime $(U, \Gamma) = (t, 0.1 t)$, while the black points correspond to strong-disorder regime $(U, \Gamma) = (t, 12 t)$. The geometries of ribbon for $\ell_\text{frac}= 10$ and $\ell_\text{frac}= 15$ are $(L, W) = (200, 48)$ and $(L, W) = (150, 72)$, respectively.  The results in topological order regime show little dependence on $\ell_{\text{frac}}$: 
the vertical separation between the two $\log M(y)$ curves corresponds to only a tiny relative difference, $\Delta M(y) \sim 10^{-3}$. The dot-dashed line represent $M(y)$ in the strongly disordered regime.The error bars are estimated from the variance of the mutual information computed over 89 (blue and red data points) and 50 (black data points) disorder realizations.}}
\label{HPG}
\end{center}
\end{figure*}

This construction reproduces the exponential stretching characteristic of the hyperbolic space, with the edges at \( y \to \pm \infty \) effectively pushed to \( y = \pm W/2 \), realizing a soft-wall geometry in the emergent curved space as illustrated in Fig.~\ref{HPG}(a).
This defines an inhomogeneous spatial geometry that effectively models a warped AdS-like background. Recall that \( x \) runs along the length of the ribbon, while \( y \) points into the bulk or embedding direction. The function \( R(y) \) captures the exponential contraction characteristic of hyperbolic space as \( |y| \) increases. It acts as a \emph{geometric shape function}, describing how far the ribbon extends in the \( x \)-direction at a given depth \( y \). Crucially, this is \emph{not} a coordinate transformation, but an embedding function that defines the intrinsic geometry of the surface.

 The corresponding horizontal geodesic distance between two boundary points at fixed depth \( y \) and horizontal separation \( \ell   \) is approximately given by 
\begin{align}
d_{\mathrm{geo}}(x_1, x_2; y) &= \int_{x_1}^{x_2} \sqrt{ds^2}=R(y)(x_2 - x_1) = R(y)\, \ell \\ 
&=\frac{W}{y_0 \cosh^2(y / y_0)} \ell . 
\label{eq:flat_geodesic} 
\end{align}
(We take $y=0$ to be the center of the ribbon.) The geodesic distance along the ribbon decays exponentially as \( y \to \infty \), reflecting the asymptotic narrowing of the geometry:
\begin{equation}
d_{\mathrm{geo}}(x_1, x_2; y) \sim \ell \cdot e^{-2y/y_0}.
\label{eq:geo_decay}
\end{equation}
This behavior is reminiscent of hyperbolic space, where distances contract or stretch exponentially with depth~\cite{weisstein2003poincare}.  
The corresponding Einstein field equations, which are provided in Appendix~\ref{EFE}, give a negative energy density, which is consistent with AdS geometry.

\section{Emergent Geodesic Distance from Coarse-Grained Mutual Information}
\label{emergent_geodesic}
Disorder induces pronounced microscopic inhomogeneities in mutual information along the 
$x$ axis, as shown in Fig.~\ref{fullMI}(a). Fractional charges appear irregularly along the zigzag edges, and their number scales linearly with ribbon length. (Conformal invariance, dependent on scale and translational symmetry, is explicitly broken.)  To probe the emergent geometry at larger scales, we coarse-grain the mutual information along the \(x\)-axis over a length scale several times larger than the average horizontal spacing between neighboring fractional charges.
Averaging over many disorder realizations serves as an effective coarse-graining procedure, since the positions of fractional charges along the edges vary randomly in each realization. The resulting averaged mutual information is independent of \(x\) but retains a dependence on \(y\) (see the definition of \(y\) below Eq.~\eqref{eq:Ry}).

In our embedded space, the coordinate \(y\) maps to the radial depth in the bulk: small \(y\) corresponds to deeper layers (infrared), whereas large \(|y|\) indicates proximity to the boundary.
Although this mapping inverts the usual AdS coordinate convention, it meaningfully reflects the physical structure of zigzag nanoribbons, where fractional charges are edge-localized.
Each pair of mirrors at \(  y \) and \( -y \) is treated as lying at the same bulk depth \( y \); the emergent geometry depends only on the radial distance, not on the sign of \( y \). In hyperbolic space, the depth is scalar: there is no distinction between \( \pm y \) in terms of curvature or geodesic distance.

{\color{black} We have computed the averaged mutual information $M(y)$ over many disorder realizations and over horizontal coordinates $x$, which results in a function of the hyperbolic distance between regions \( I \) and \( J \), see Fig.~\ref{HPG}(b). 
It is well described by the following expression:}
% We have computed the mutual information averaged with disorder \( M(y) \) as a function of the hyperbolic distance between regions \( I \) and \( J \), see Fig.~\ref{HPG}(b). 
% The resulting mutual information \( M(y) \) is well described a by the following expression
\begin{equation}
    M(y) \propto \left\langle M(x_1, x_2, y) \right\rangle 
    \approx A e^{-\alpha \, d_{\mathrm{geo}}(y)},
    \label{eq:MI_decay}
\end{equation}
where the effective hyperbolic distance \( d_{\mathrm{geo}}(y) \) is given by Eq.~\eqref{eq:flat_geodesic} with \( \ell \) being replaced by \( \ell_{\text{frac}} \), which characterizes the typical spatial distance between fractional charges, and
$\alpha$ is a certain positive constant (see the caption of Fig.~\ref{HPG}(b) for a particular value of $\alpha$ and $A$).
\textit{This coarse-grained mutual information defines a geodesic distance in the information geometry - an emergent feature shaped by correlations rather than by explicit spatial embedding.}
The coarse-grained mutual information \( M(y) \) captures the nonlocal entanglement between fractional charges. The exponential decay along the $y$-axis mirrors the behavior of correlation functions in holographic settings such as AdS/CFT, where geodesic lengths or minimal surfaces determine the strength of correlations~\cite{Ryu2006} (see Appendix~\ref{distance}). This justifies interpreting the coarse-grained, disorder-averaged system as realizing an emergent hyperbolic-like geometry.  \textcolor{black}{Note that, in the absence of semions or in the strong disorder regime, the geometry is qualitatively different. In particular, the AdS geometry does not emerge, since the exponential behavior fails to develop; see Fig.~\ref{HPG}(b).} {\color{black}In the parameter regime where the topological entanglement entropy is universal, we find that the constant $\alpha$ is also universal. Within numerical accuracy, it exhibits no dependence on the disorder strength $\Gamma$ or the on-site interaction $U$.}

% \begin{figure}[h!]
%     \centering
%     \includegraphics[width=0.5\linewidth]{figure10.pdf}
%     \caption{The coarse-grained (disorder-averaged) mutual information 
% $\log M(y)$ is computed with a transverse width $w = 1$, 
% for typical separations between fractional charges 
% $\ell_{\text{frac}} = 10,\,15$.  The fitted slope yields an inverse decay length 
% $\alpha = 0.09$ (see Eq.~(\ref{eq:MI_decay})), with the curvature scale parameter fixed at $y_0 = 10$. The five data points correspond to vertical separations \( \Delta y = 45, 43, 41, 39 \) between \( I \) and \( J \), with \( |y| = \Delta y / 2 \). Parameters for Hamiltonian are $(U, \Gamma) = (t, 0.1 t)$. The geometries of ribbon for $\ell_\text{frac}= 10$ and $\ell_\text{frac}= 15$ are $(L, W) = (200, 48)$ and $(L, W) = (150, 72)$, respectively. The results show little dependence on $\ell_{\text{frac}}$: 
% the vertical separation between the two $\log M(y)$ curves corresponds 
% to only a tiny relative difference,
% $\Delta M(y) \sim 10^{-3}$.}
%     \label{fig:mutual_hyperdistance1}
% \end{figure}

Our results suggest that the emergent space reconstructed from the entanglement correlations behaves as if it is hyperbolic. This perspective aligns with Swingle’s proposal~\cite{Swingle2012} and the broader entanglement = geometry program initiated by Van Raamsdonk~\cite{van2010}. As in AdS/CFT, the decay of correlations with increasing geodesic length reflects an underlying emergent geometry. Here, the bulk geometry arises not from fundamental curvature but from the entanglement structure itself, revealing a spatial slice of a higher-dimensional curved spacetime emergent from quantum information.

% \pagebreak

\section{Summary}
\label{summary}

In this work, we examined the highly entangled (strongly correlated) ground state of ZGNRs and shown that their behavior can be effectively described in terms of an emergent space-time geometry. In particular, we identified the regimes and conditions under which this emergent geometry remains robust, providing criteria for its stability and persistence. A key diagnostic is the mutual information between anyon fractional charges located on opposite zigzag edges, which serves as a sensitive probe of the underlying entanglement structure. Its spatial profile encodes an effective bulk metric 
\( g_{\mu\nu}(x,y) \), thereby providing a coarse-grained 
mapping from the underlying microscopic lattice to an 
emergent geometry reminiscent of hyperbolic space. In this description, edge-to-edge mutual information plays the role of a geodesic distance in a negatively curved space, reminiscent of the AdS/CFT correspondence. The scale-dependent decay of mutual information thus substantiates the emergence of an AdS-like bulk geometry, offering a concrete realization of holography in a system that lacks conformal symmetry or intrinsic curvature, yet still manifests the deep connection between entanglement and geometry. 

\textcolor{black}{Anyonic phenomena in quasi-one-dimensional systems provide an instructive contrast to those found in genuinely two-dimensional topological phases. 
In particular, two-leg electron ladders hosting Abelian anyons belong to the same universality class as zigzag graphene nanoribbons~\cite{Yang2025} and share a character that cannot be classified within the framework of modular tensor categories (MTCs)~\cite{Kitaev2006/2}, in sharp contrast to intrinsically topologically ordered two-dimensional systems.  In (2+1)-dimensional gapped phases with intrinsic topological order, quasiparticle excitations obey fractional statistics and possess fusion and braiding properties that satisfy the axioms of a modular tensor category; hence their excitations are naturally described by an MTC.
Related anyon-inspired properties have also been explored in other one-dimensional platforms, including Rydberg atom lattices~\cite{Lesanovsky2012}
and supersymmetric lattice models~\cite{Jana2023}, where the Hilbert space is
constrained to realize the fusion algebra of non-Abelian anyons. While these
systems do not support genuine anyonic braiding or intrinsic topological order, they nevertheless exhibit nontrivial entanglement structures, highlighting the richness of anyon-inspired physics beyond strictly two-dimensional settings.}

\textcolor{black}{Traditionally, anyons have been classified primarily through their quantum statistical phases, namely their exchange properties. In contrast, our approach examines semions from an information-theoretic perspective, focusing on the structure of the hidden ``spacetime'' encoded in their entanglement patterns. This viewpoint naturally raises several intriguing questions. Can anyons be systematically classified according to the information-theoretic geometries they generate? Do different types of anyons give rise to distinct emergent geometrical structures?  Disordered zigzag graphene nanoribbons exhibit Abelian semionic structure already at the level of the ground state. In contrast, in physical realizations described by MTCs, anyons, potentially including non-Abelian species, arise as quasiparticle excitations above the ground state. Since the computation of entanglement for excited states {\color{black} and for systems supportings non-Abelian anyons} is subtle and lacks a general framework~\cite{Balachandran2013}, an important open question is how features typically associated with MTC descriptions are encoded in the emergent entanglement geometry. Addressing this question across different intrinsically topologically ordered phases may provide a new information-theoretic route to understanding topological phases of matter.}

Data supporting the findings of this article are openly available~\cite{le_2025_16916229}. {\color{black}Numerical calculations were performed using HFGN code~\cite{Le_graphene_ribbon_HartreeFock}.}

%TC:ignore
\acknowledgments
H.-A.L acknowledges the financial support from the Institute for Basic Science (IBS) under grant IBS-R027-D1. Numerical calculations were performed using the GORDON GPU clusters of the IBS Center for Quantum Nanoscience.

%%%%%%%%%%%%%%%%%%%%%%%%%%%%%%%%%%%%%%%%%%%%%%%%%%%%%%%%%%

%%%%%%%%%%%%%%%%%%%%%%%%%%%%%%%%%%%%%%%%%%%%%%%%%%%%%%%%%%
% \section*{References}
% \bibliographystyle{apsrev4-2}
% \bibliographystyle{iopart-num}
\bibliography{aapmsamp}
%TC:endignore

\appendix

\section{Gaussian Curvature of the Induced Metric}\label{GCurv}

The general form of such a two-dimensional warped metric is:
\begin{equation}
ds^2 = f(y)^2 dx^2 + dy^2,
\label{eq:warped_metric}
\end{equation}
which explicitly encodes how distances along the \(x\)-direction vary with position in \(y\).
{\color{black} The metric} has Gaussian curvature given by
\begin{equation}
K = -\frac{f''(y)}{f(y)}.
\label{eq:gaussian_curvature}
\end{equation}
{\color{black}
In our case, $f(y) = W/[y_0 \cosh^2(y/y_0)]$ so the curvature takes the following form
\begin{equation}
K = \frac{2}{y_0^2} \left(\frac{3}{\cosh^2{(y/y_0)}} - 2 \right).
\label{eq:scale_factor}
\end{equation}}
The curvature is negative for large $|y|$.  The geometry is hyperbolic-like near the edges. This confirms that the induced metric describes a space with constant negative curvature at small scales, consistent with the qualitative features of Anti-de Sitter (AdS) geometry. Such curvature arises purely from the scale-dependent entanglement pattern encoded in the ground-state wavefunction, rather than from any gravitational dynamics.

\section{Einstein Field Equations}\label{EFE}

While the physical ribbon remains embedded in a flat background spacetime, the nonlocal correlations in the quantum ground state define an effective hyperbolic-like geometry. This emergent geometry is not a consequence of gravitational dynamics, but a reflection of the scale-dependent entanglement encoded in the ground-state wavefunction.  Let's find the energy-momentum tensor of Einstein's field equation.

We first find the Riemann curvature tensor from the metric~\cite{Susskind2023}.
Consider a metric of the following form:
\be
\label{metric1}
d s^2 = - d t^2 + f(y)^2 d x^2 + d y^2.
\ee
The metric (\ref{metric1}) is diagonal and static. Its non-trivial component $g_{xx}$ depends only on $y$, and these features simplify the calculation of the curvature tensor considerably.

The affine connection~\cite{zee2013einstein} is given by
\be
\label{affine-connection}
\Gamma^{\lambda}_{\mu \nu} = \frac{1}{2} g^{\lambda \sigma}
\Big( \partial_\mu g_{\sigma \nu} +  \partial_\nu g_{\sigma \mu} -
\partial_\sigma g_{\mu \nu} \Big ).
\ee
If $\lambda = t$, then owing to the diagonal metric we have $\sigma = \lambda = t$. 
Since the metric is static, we easily see that 
$\Gamma^t_{\mu \nu} = 0$.

If $\lambda = x$, again $ \sigma = \lambda = x$. Then, the last term of Eq.~\eqref{affine-connection} drops out,
and $\nu =x, \mu =y$ contributes. Taking into account the symmetry of the affine connection, we can obtain
$\Gamma^x_{yx} = \Gamma^x_{xy} = f'(y)/f(y)$. 

Lastly, if $\lambda = y$, $ \sigma = \lambda = y$.
The last term of Eq.~(\ref{affine-connection}) gives $-f'(y) f(y)$ for $\mu=\nu =x$, while the first 
two terms give vanishing result since all metric components involving $y$ index are constants. We have
\be
\Gamma^x_{x y}=\Gamma^x_{y x} =  f'(y)/f(y), \quad 
\Gamma^y_{xx} = -f'(y) f(y),
\ee
and all other components vanish.

Next the Riemann curvature tensor is given by
\be
R^\lambda{}_{\mu \nu \sigma} =\partial_\nu \Gamma^{\lambda}_{\mu \sigma} - 
\partial_\sigma \Gamma^{\lambda}_{\mu \nu} +\Gamma^\eta_{\mu \sigma} \Gamma^\lambda_{\eta  \nu}
-\Gamma^\eta_{\mu \nu} \Gamma^\lambda_{\eta  \sigma},
\ee
where the anti-symmetry under $\nu \longleftrightarrow \sigma$ is manifest.

Clearly, if $\lambda = t$, the curvature tensor vanishes trivially. And it is obvious that all other indices can be only $x$ or $y$
for non-zero results.
Start with $\lambda = x, \mu = x$, $R^x{}_{x \nu \sigma}$. Due to anti-symmetry of $\nu \sigma$ there is 
only one possibility: $ \nu =x, \sigma =y$. An easy calculation shows that $R^x{}_{x \nu \sigma}=0$
(note crucial cancellation between derivative and nonlinear terms).
As of $\lambda = x, \mu =y$, the similar argument shows that the nontrivial component is 
$R^x{}_{y x y}=-f^{\prime \prime}(y)/f(y) = -R^x{}_{y y x} $.
The case $\lambda =y, \mu = x$ can be treated identically, yielding
$R^y{}_{x x y}=+f^{\prime \prime}(y) f(y) = -R^y{}_{x y x} $.
For the case of $\lambda =y, \mu = y$, all relevant components of affine connection vanish, so does 
the Riemann curvature tensor.
Summarizing, the non-vanishing components of Riemann curvature tensor are 
\begin{align}
R^x{}_{y x y}&=-f^{\prime \prime}(y)/f(y) = -R^x{}_{y y x}, \\
R^y{}_{x x y}&=+f^{\prime \prime}(y) f(y) = -R^y{}_{x y x}.
\end{align}
From the above result, the Ricci tensor \(R_{\mu \sigma} = R^{\lambda}{}_{\mu \lambda \sigma}\) can be read off. The non-vanishing components are
\begin{equation}
R_{xx} = -f''(y) f(y), \quad 
R_{yy} = -\frac{f''(y)}{f(y)}.
\end{equation}
The Ricci scalar is 
\be 
R=R^\mu{}_\mu =g^{\mu \nu} R_{\mu \nu} =  -2 f^{\prime \prime}(y)/f(y).
\ee
%All of the above analytic results can be checked readily using a symbolic calculation program such as Mathematica.
The only non-vanishing component of the Einstein tensor $G_{\mu \nu} = R_{\mu \nu} - \frac{1}{2} g_{\mu \nu} R$ is (all spatial components are cancelled out)
\be
G_{tt} = -f^{\prime \prime}(y)/f(y).
\ee
The Einstein equation 
$ G_{\mu \nu} = \frac{8\pi G}{c^4} T_{\mu \nu}$ gives the corresponding energy momentum tensor
\be 
T_{tt}= -\frac{c^4}{8 \pi G} \, \frac{f^{\prime \prime}(y)}{f(y)},
\ee 
and all other components vanish.   This implies that the asymptotic region is supported by a negative energy density, which commonly arises in AdS-like warped geometries to sustain their curvature.

\section{Correlations in the Hyperbolic Space}\label{distance}

A key feature of conformal field theories (CFTs) describing critical systems in flat (Euclidean) space is the algebraic decay of two-point correlators \cite{Cardybook}:
\begin{equation}
\langle \mathcal{O}(x) \mathcal{O}(0) \rangle \sim \frac{1}{|x|^{2\alpha}}.
\end{equation}

In the AdS/CFT correspondence, the boundary theory is conformally invariant, so this same functional form arises \cite{Witten1998, Hartnoll2009}. In the AdS bulk, by contrast, the geodesic distance between two boundary points grows only logarithmically with their coordinate separation at  large distances:
\begin{equation}
d_{\mathrm{geo}}(x_1, x_2) \sim \log\left( \frac{|x_1 - x_2|}{\epsilon} \right),
\end{equation}
where $\epsilon$ is a UV cutoff near the boundary (with AdS length scale set to $R = 1$).
As a result, the algebraic decay of the boundary becomes exponential in terms of the bulk geodesic distance:
\begin{equation}
\langle \mathcal{O}(x_1), \mathcal{O}(x_2) \rangle \sim e^{-\alpha d_{\mathrm{geo}}(x_1, x_2)}.
\end{equation}
Therefore, exponential decay of mutual information supports the interpretation of the system as residing in an emergent curved space with geodesic-like correlation structure.
  This encapsulates a central idea of AdS/CFT: correlation functions and entanglement on the boundary encode geometric distances in the emergent hyperbolic bulk.

\end{document}